\documentclass[12pt]{article}
\usepackage{a4,graphicx,amssymb}
\usepackage[small,bf]{caption}

\newcommand{\mnote}[1]{}                   

\newcommand{\half}{\mbox{\small $\frac{1}{2}$}}          
\newcommand{\quart}{\mbox{\small $\frac{1}{4}$}}         
\newcommand{\third}{\mbox{\small $\frac{1}{3}$}}         
\newcommand{\twothird}{\mbox{\small $\frac{2}{3}$}}      
\newcommand{\fourthird}{\mbox{\small $\frac{4}{3}$}}     
\newcommand{\msbar}{\mbox{\tiny $\overline{MS}$}}        
\newcommand{\rgi}{\mbox{\tiny $R\!G\!I$}}                
\newcommand{\bare}{\mbox{\tiny $bare$}}                  
\newcommand{\lat}{\mbox{\tiny $lat$}}                    
\newcommand{\plaq}{\Box}                                 

\def\lsim{\mathrel{\rlap{\lower4pt\hbox{\hskip1pt$\sim$}}
    \raise1pt\hbox{$<$}}}                
\def\gsim{\mathrel{\rlap{\lower4pt\hbox{\hskip1pt$\sim$}}
    \raise1pt\hbox{$>$}}}                

\begin{document}

\title{
\vspace{-3.0cm}
\flushright{\normalsize ADP-12-24/T791} \\
\vspace{-0.35cm}
{\normalsize DESY 12-079} \\
\vspace{-0.35cm}
{\normalsize Edinburgh 2012/06} \\
\vspace{-0.35cm}
{\normalsize Liverpool LTH 945} \\
\vspace{-0.35cm}
{\normalsize July 30, 2012} \\
\vspace{0.5cm}
\centering{\Large \bf A Lattice Study of the Glue in the Nucleon}}

\author{\large
        R. Horsley$^a$, R. Millo$^b$, Y. Nakamura$^c$, H. Perlt$^d$, \\
        D. Pleiter$^{ef}$, P.~E.~L. Rakow$^b$, G. Schierholz$^{fg}$, \\
        A. Schiller$^d$, F. Winter$^a$ and J.~M. Zanotti$^h$ \\[1em]
         -- QCDSF-UKQCD Collaboration -- \\[1em]
        \small $^a$ School of Physics and Astronomy,
               University of Edinburgh, \\[-0.5em]
        \small Edinburgh EH9 3JZ, UK \\[0.25em]
        \small $^b$ Theoretical Physics Division,
               Department of Mathematical Sciences, \\[-0.5em]
        \small University of Liverpool,
               Liverpool L69 3BX, UK \\[0.25em]
        \small $^c$ RIKEN Advanced Institute for
               Computational Science, \\[-0.5em]
        \small Kobe, Hyogo 650-0047, Japan \\[0.25em]
        \small $^d$ Institut f\"ur Theoretische Physik,
               Universit\"at Leipzig, \\[-0.5em]
        \small 04109 Leipzig, Germany \\[0.25em]
        \small $^e$ J\"ulich Supercomputer Centre,
               Forschungszentrum J\"ulich, \\[-0.5em]
        \small 52425 J\"ulich, Germany \\[0.25em]
        \small $^f$ Institut f\"ur Theoretische Physik,
               Universit\"at Regensburg, \\[-0.5em]
        \small 93040 Regensburg, Germany \\[0.25em]
        \small $^g$ Deutsches Elektronen-Synchrotron DESY, \\[-0.5em]
        \small 22603 Hamburg, Germany \\[0.25em]
        \small $^h$ CSSM, School of Chemistry and Physics,
               University of Adelaide, \\[-0.5em]
        \small Adelaide SA 5005, Australia}

\date{}

\maketitle



\begin{abstract}
By introducing an additional operator into the action and using the
Feynman--Hellmann theorem we describe a method to determine
both the quark line connected and disconnected terms of matrix elements.
As an illustration of the method we calculate the gluon contribution
(chromo-electric and chromo-magnetic components) to the nucleon mass.
\end{abstract}


\clearpage


\section{Introduction} 
\label{intro}


One of the earliest experimental indications that the nucleon consists
not only of three quarks, but also has a gluonic contribution came
from the measurement of the fraction of the nucleon momentum
carried by the quarks. That this did not sum up to $1$ as is required
from the energy--momentum sum rule gave evidence for the
existence of the gluon. Denoting $\langle x \rangle_f$ as the
fraction of the nucleon momentum carried by parton $f$ we have
\begin{eqnarray}
   \sum_q \langle x \rangle_q + \langle x \rangle_g = 1 \,,
\label{intro.a}
\end{eqnarray}
where for the quarks $f \equiv q = u, d, \ldots$ and for the gluon
$f \equiv g$. Experimentally $\langle x \rangle_{u+d} \sim 0.4$ so the
missing component is large $ \sim 50\%$ of the total nucleon
momentum. Both $\langle x \rangle_q$ and $\langle x \rangle_g$
have similar definitions and so analogously to the definition of
$\langle x \rangle_q$ we have, with ${\cal M}$ denoting Minkowski space
\begin{eqnarray}
   \langle N(\vec{p}) | [ \widehat{\cal O}^{{\cal M}(g)\mu_1\mu_2} -
              \quart \eta^{\mu_1\mu_2} \widehat{\cal O}^{{\cal M}(g)\alpha}_{
                                     \phantom{{\cal M}(g)\alpha}\alpha}
                        ] | N(\vec{p}) \rangle
    = 2 \langle x \rangle_g \left[ p^{\mu_1}p^{\mu_2} -
                                       \quart \eta^{\mu_1\mu_2} m_N^2
                            \right] \,,
\label{intro.b}
\end{eqnarray}
where
\begin{eqnarray}
   O^{{\cal M}(g)\mu_1\mu_2}
      = - \mbox{tr}_c F^{{\cal M}\mu_1\alpha}
                    F^{{\cal M}\mu_2}_{\phantom{{\cal M}\mu_2}\alpha} \,,
\end{eqnarray}
(where ${\cal O}(t) = \int d^3x\,O(t,\vec{x})$ and with normalisation
$\langle N(\vec{p})|N(\vec{p}^{\,\prime})\rangle 
= 2E_N\delta(\vec{p}-\vec{p}^{\,\prime})$).
Note that we can generalise from a nucleon to an arbitrary hadron
(averaging over polarisations if necessary). Higher moments
can also be considered, by inserting covariant derivatives
between the $F$s. These occur when using the Wilson operator
product expansion which relates them to moments
of structure functions in a twist expansion.

There have been many lattice estimates of the quark momentum fraction
$\langle x \rangle_q$ both for the nucleon (see e.g.\ 
\cite{renner10a,hagler09a} for a review) and the pion e.g.\
\cite{best97a,guagnelli04a}, but few attempts for the gluon part,
$\langle x \rangle_g$ \cite{gockeler96b,meyer07a,liu12a}. This is due 
to the fact that a lattice simulation must compute a quark line
disconnected term, which is extremely noisy and gives a poor signal.
These are direct calculations; in this letter we propose a new method
using the Feynman-Hellmann theorem, to determine the gradient of $E_N$ as
a function of a parameter of an operator which has been introduced
into the action $S \to S(\lambda) = S + \lambda S_O$.
An obvious disadvantage of this method is that it requires dedicated
simulations for each operator of interest, but the gain,
as we shall see, is a much cleaner signal.

While the method is general, we shall demonstrate its practicability
here by determining $\langle x \rangle_g$ in the quenched case.


\section{The Feynman--Hellmann theorem}
\label{FNsection}


We first briefly describe the Feynman--Hellmann theorem, in a Euclidean form
that will be useful for the case to be considered here. Let $S$ depend
on some parameter $\lambda$, so $S \to S(\lambda)$. Now as by definition
the (Euclidean) correlation function is given by
\begin{eqnarray}
   \langle N(t) \overline{N}(0) \rangle_\lambda 
      \equiv { \int [dU]N(t)\overline{N}(0) e^{-S(\lambda)} \over
                  \int [dU] e^{-S(\lambda)} } \,,
\end{eqnarray}
(the unpolarised case for the nucleon and where we make the obvious
replacements $N$ by $H$ and $\overline{N}$ by $H^\dagger$ for other
hadrons), then we have
\begin{eqnarray}
   {\partial \over \partial \lambda} \langle N(t)\overline{N}(0) \rangle_\lambda
      = - \left\langle N(t)\left(
            {\partial S(\lambda) \over \partial\lambda}
               - \langle {\partial S(\lambda) \over \partial\lambda}
                                                          \rangle_\lambda
                      \right) \overline{N}(0) \right\rangle_\lambda \,.
\label{intermediate_fh}
\end{eqnarray}
We now use the transfer matrix formalism on both sides of this equation.
Ignoring finite size effects this gives
\begin{eqnarray}
   \langle N(t) \overline{N}(0) \rangle_\lambda 
      = A_N(\lambda)e^{-E_N(\lambda)t} + \mbox{exp. smaller terms} \,.
\end{eqnarray}
so on the LHS of eq.~(\ref{intermediate_fh}),
\begin{eqnarray}
   {\partial \over \partial \lambda}
                  \langle N(t)\overline{N}(0) \rangle_\lambda
      = - {\partial E_N(\lambda) \over \partial \lambda}
               \langle N(t)\overline{N}(0) \rangle_\lambda\,t
        + \mbox{exp. smaller terms} \,.
\label{lhs_fh}
\end{eqnarray}
Furthermore, if $\Omega(\tau)$ is any operator (local in time), then 
using the transfer matrix formalism again the associated $3$-point
function gives
\begin{eqnarray}
   { \langle N(t) \Omega(\tau)\overline{N}(0) \rangle_\lambda
     \over \langle N(t)\overline{N}(0) \rangle_\lambda }
   =  \left\{ 
         \begin{array}{ll}
            {1 \over 2E_N(\lambda)} 
               \langle N|\widehat{\Omega}|N\rangle_{\lambda}
                + \mbox{exp. small terms} & 0 \ll \tau \ll t  \\[0.1in]
            \mbox{exp. small terms}       & \mbox{otherwise}  \\
               \end{array}
      \right. \,.
\end{eqnarray}
Note that we have inserted a $2E_N$ in the denominator of the RHS
to account for the mis-match of normalisations, i.e.\ to agree
with those of eq.~(\ref{intro.b}). Hence summing over $\tau$
also gives a linear term in $t$.
Thus from this equation, replacing $\sum_\tau \Omega(\tau)$ 
by the operator in the RHS of eq.~(\ref{intermediate_fh}),
and together with eq.~(\ref{lhs_fh}) we have the
Feynman--Hellmann theorem
\begin{eqnarray}
   {\partial E_N(\lambda) \over \partial \lambda}
      = {1 \over 2E_N(\lambda)}
        \left\langle N \left|
              : \widehat{\partial S(\lambda) \over \partial\lambda} :
                       \right|N \right\rangle_\lambda \,,
\label{fh_thm}
\end{eqnarray}
(where $:\ldots:$ means that the vacuum term has been subtracted).
Thus by suitably choosing $S_O$ and by identifying numerically the gradient
of $E_N(\lambda)$ at $\lambda = 0$ we can determine the desired matrix element.


\section{The lattice method}
\label{lattice}


\subsection{Gluon operators}
\label{gluon_ops}


Before considering the lattice, let us first Euclideanise the
gluon operators%
\footnote{Our conventions follow \cite{best97a}. So
$E^{{\cal M}i} = F^{{\cal M}i0} \to iF_{i4} \equiv iE_i$ and
$B^{{\cal M}i} = -{\half}\epsilon^{ijk}F^{\cal M}_{\phantom{\cal M}jk} \to
{\half}\epsilon_{ijk}F_{jk} \equiv B_i$. \label{foot_euclid}}
to give us an indication of what we might add to the action. Defining
\begin{eqnarray}
   O_{\mu\nu} = -\mbox{tr}_cF_{\mu\alpha} F_{\nu\alpha} \,,
\end{eqnarray}
($\mbox{tr}_c F^2 = \half F^{a\,2}$) this then gives the two
obvious operator choices $(a)$ and $(b)$,
\begin{eqnarray}
   O_{a\,i}
       &=& O_{i4} 
            = \mbox{tr}_c( \vec{E} \times \vec{B} )_i
                                       \nonumber  \\
   O_b &=& O_{44} - {\third} O_{jj}
            = {\twothird} \mbox{tr}_c( -\vec{E}^2 + \vec{B}^2 )
\label{lattice.a}
\end{eqnarray}
($O^{{\cal M}(g)}_a \to i O_a$ and $O^{{\cal M}(g)}_b \to O_b$).
The relation to $\langle x \rangle_g$ is given by
\begin{eqnarray}
   \langle N(\vec{p}) | \widehat{\cal O}_{a\,i} | N(\vec{p}) \rangle 
         &=& - 2 i E_N p_i \, \langle x \rangle_g
                                       \nonumber  \\
   \langle N(\vec{p}) | \widehat{\cal O}_b | N(\vec{p}) \rangle 
         &=& 2 (m_N^2 + \fourthird\vec{p}^{\,2}) \, \langle x \rangle_g \,,
\label{lattice.b}
\end{eqnarray}
with
\begin{eqnarray}
   \widehat{\cal O}_{a\,i} 
      = \mbox{tr}_c (\vec{\widehat{\cal E}}\times\vec{\widehat{\cal B}})_i
                                               \,, \qquad
   \widehat{\cal O}_b
      = \twothird\mbox{tr}_c (-\vec{\widehat{\cal E}}^2 
                               + \vec{\widehat{\cal B}}^2) \,.
\end{eqnarray}
Both choices have their difficulties: operator $(a)$ always needs a non-zero
momentum $\vec{p}$, while operator $(b)$ requires a delicate subtraction
between two terms similar in magnitude.

Note that, because of Euclideanisation (footnote~\ref{foot_euclid})
the {\it energy} has a negative ${\cal E}^2$ term, while the {\it action}
(see section~\ref{action}) has a positive ${\cal E}^2$ term.


\subsection{The action}
\label{action}


We now turn to the lattice. We shall use the Wilson gluonic action
\begin{eqnarray}
   S = \third\beta \sum_{x\, \mu < \nu}
           \mbox{Re}\,\mbox{tr}_c
           \left[ 1 - U_{\mu\nu}^{\plaq}(x) \right] \,,
\label{Wil_glue_act}
\end{eqnarray}
(i.e.\ sum over plaquettes), with $\beta = 6 /g^2$.
As 
\begin{eqnarray}
   \mbox{Re}\,\mbox{tr}_c \left[1 - U_{\mu\nu}^{\plaq}(x) \right]
         = \quart a^4 g^2 F_{\mu\nu}^a(x)^2 + \ldots \,,
\end{eqnarray}
this motivates the simplest
definition of electric and magnetic field on each time slice as
\begin{eqnarray}
   \half{\cal E}^{a\,2}(\tau)
      &=& \third\beta \frac{1}{a} \sum_{\vec{x}\,i}\,\mbox{Re}\,\mbox{tr}_c
               \left[1 - U_{i4}^{\plaq}(\vec{x},\tau) \right]
                                                         \nonumber \\
   \half{\cal B}^{a\,2}(\tau)
      &=& \third\beta \frac{1}{a} \sum_{\vec{x}\,i<j}\, \mbox{Re}\,\mbox{tr}_c
               \left[1 - U_{ij}^{\plaq}(\vec{x},\tau) \right] \,,
\label{EBdef}
\end{eqnarray} 
respectively. For the action we thus take
\begin{eqnarray}
   S(\lambda)
     = a \sum_\tau\half[{\cal E}^{a\,2}(\tau) + {\cal B}^{a\,2}(\tau)]
          - \lambda a \sum_\tau\half[-{\cal E}^{a\,2}(\tau)
                                  + {\cal B}^{a\,2}(\tau)] \,,
\end{eqnarray} 
or in terms of the gauge plaquettes
\begin{eqnarray}
   S(\lambda)
      &=& \third\beta(1+\lambda) \sum_i\,\mbox{Re}\,\mbox{tr}_c 
               \left[1 - U_{i4}^{\plaq}(\vec{x},\tau) \right]
                                                         \nonumber \\ 
      & & + \third\beta(1-\lambda) \sum_{i<j}\, \mbox{Re}\,\mbox{tr}_c 
               \left[1 - U_{ij}^{\plaq}(\vec{x},\tau) \right] \,.
\label{S_lambda}
\end{eqnarray}
Of course for $\lambda = 0$, then this reduces to the standard action,
eq.~(\ref{Wil_glue_act}).


\subsection{Gluon moment}


Comparing the results of sections~\ref{gluon_ops} and \ref{action}
we see that they can be applied to operator (b) only; operator (a)
would require the clover definition of the field strength tensor.
Using eq.~(\ref{lattice.a}) together with eq.~(\ref{lattice.b})
and eq.~(\ref{fh_thm}) gives from the Feynman--Hellmann theorem
\begin{eqnarray}
   { \partial E_N(\lambda) \over \partial \lambda }
      = - {1 \over 2E_N(\lambda)}\,
           \langle N(\vec{p})| \half(-\widehat{\cal E}^{a\,2} 
                                +\widehat{\cal B}^{a\,2})
                             | N(\vec{p}) \rangle_\lambda \,,
\end{eqnarray}
which leads to
\begin{eqnarray}
   \left. {\partial E_N(\lambda)\over \partial \lambda} \right|_{\lambda = 0}
      = - {3 \over 2E_N} \, \left( m_N^2 + \fourthird\vec{p}^{\,2} \right)
                             \langle x \rangle_g^{\lat} \,,
\label{FH_xg}
\end{eqnarray}
where the $^{\lat}$ superscript on $\langle x \rangle_g^{\lat}$ signifies
that it is now the lattice operator.

The vacuum term which appears in section~\ref{FNsection} has been
dropped, because
\begin{eqnarray}
   \langle 0 | \half(-\widehat{\cal E}^{a\,2}
              +\widehat{\cal B}^{a\,2}) |0 \rangle = 0  \,.
\label{vacterm}
\end{eqnarray}
This follows from rotation symmetry. In the Euclidean
vacuum the time and space  directions are equivalent, so
the average trace of the chromo-electric plaquettes, $U_{i4}^\plaq$,
is the same as that of the chromo-magnetic plaquettes, $U_{ij}^\plaq$, in
eq.~(\ref{EBdef}), leading to perfect cancellation in eq.~(\ref{vacterm}).


\section{Lattice results}
\label{lat_results}


We work with quenched Wilson clover fermions at $\beta=6.0$, $c_{sw} = 1.769$
and $\kappa = 0.1320$, $0.1324$, $0.1333$, $0.1338$, $0.1342$
on a $24^3\times 48$ lattice with antiperiodic time boundary conditions
for the fermion.
We have generated
O(500) configurations for each ensemble. We use standard nucleon
interpolating operators together with Jacobi smeared source/sink as in 
e.g.\ \cite{best97a}. The results were generated using the Chroma
program suite, \cite{edwards04a}. We have only considered the case
$\vec{p} = \vec{0}$ so eq.~(\ref{FH_xg}) reduces to
\begin{eqnarray}
   \langle x \rangle_g^{\lat}
      =  - {2 \over 3am_N}\,
         \left. {\partial am_N(\lambda)\over \partial \lambda} 
         \right|_{\lambda = 0} \,.
\label{xg_bare}
\end{eqnarray}
To estimate the gradient at $\lambda = 0$, we have generated data at
$\lambda = -0.03333$, $0.0$, $0.03333$ which enables us to straddle
the $\lambda = 0$ point. The raw data results are given in
Table~\ref{raw_res}.
\begin{table}[htb]
   \begin{center}
      \begin{tabular}{c|ccc}
         $\kappa$ & $\lambda=-0.03333$ & $\lambda = 0$ & $\lambda=0.03333$ \\
         \hline
         0.1320  & 1.0033(29) & 0.9772(33) & 0.9564(34) \\
         0.1324  & 0.9537(30) & 0.9283(34) & 0.9077(36) \\
         0.1333  & 0.8357(33) & 0.8117(40) & 0.7923(41) \\
         0.1338  & 0.7649(38) & 0.7413(47) & 0.7236(47) \\
         0.1342  & 0.7044(47) & 0.6799(62) & 0.6647(55) \\
         \hline
      \end{tabular}
   \end{center}
\caption{Nucleon masses, $am_N$, as a function of $\lambda$
         for five quark masses, $\kappa$, calculated on
         ensembles with fixed $\beta = 6.0$ and $c_{sw}=1.769$.}
\label{raw_res}
\end{table}

\begin{figure}[htbp]
   \begin{center}
      \includegraphics[width=10.00cm]
         {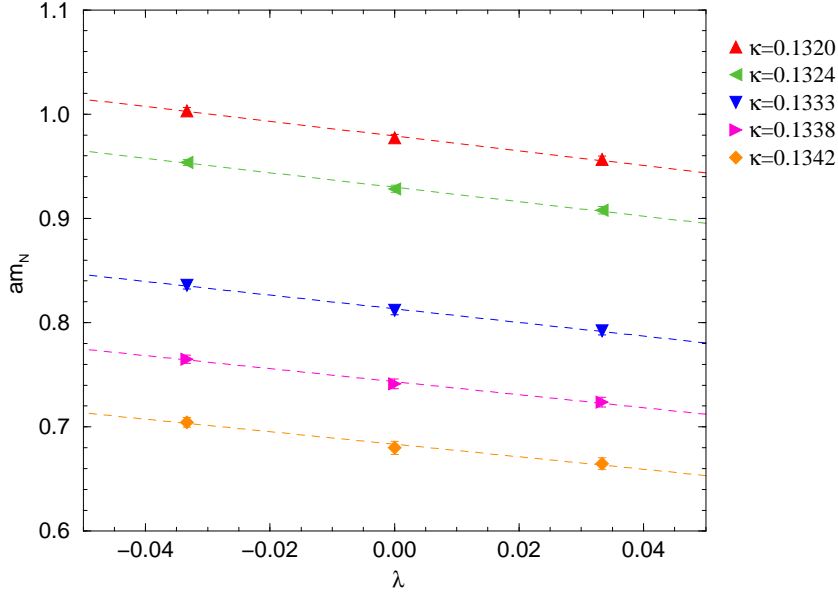}
   \end{center}
\caption{The nucleon mass against $\lambda$ for the five $\kappa$ values,
         together with a linear fit for each $\kappa$ value.}
\label{la_mn}
\end{figure}

In Fig.~\ref{la_mn} we plot the nucleon mass, $am_N$, against $\lambda$
for the five quark masses. The data show no $O(\lambda^2)$ effects
for the $\lambda$ values chosen. These gradients (at $\lambda = 0$)
together with the nucleon masses (again at $\lambda = 0$) determine
$\langle x \rangle_g^{\lat}$ from eq.~(\ref{xg_bare}) which
are given in Table~\ref{xg_lat_res}.
\begin{table}[htb]
   \begin{center}
      \begin{tabular}{c|cc}
         $\kappa$ & $am_\pi$    & $\langle x \rangle_g^{\lat}$  \\
         \hline
         0.1320  & 0.55499(48) & 0.4826(456) \\
         0.1324  & 0.51745(49) & 0.4985(502) \\
         0.1333  & 0.42531(52) & 0.5383(644) \\
         0.1338  & 0.36711(55) & 0.5620(811) \\
         0.1342  & 0.31433(62) & 0.5893(1062)\\
         \hline
      \end{tabular}
   \end{center}
\caption{The pion mass and $\langle x \rangle_g^{\lat}$ for the five
         different quark masses.}
\label{xg_lat_res}
\end{table}


\section{Renormalisation}
\label{renormalisation}


As gluon operators are singlets, they can mix with the quark singlet.
However there exists a combination of singlet operators with vanishing
anomalous dimension. (This is due to the conservation of the
energy-momentum tensor, eq.~(\ref{intro.a}).) We follow \cite{meyer07a}
and first write
\begin{eqnarray}
    \langle x \rangle_g^{\bare} 
            + \sum_q \langle x \rangle_q^{\bare} = 1 + O(a^2) \,,
\label{bare_sum_rule}
\end{eqnarray}
where
\begin{eqnarray}
   \langle x \rangle_g^{\bare} = Z_g  \langle x \rangle_g^{\lat} \,,
   \qquad
   \langle x \rangle_q^{\bare} = Z_q  \langle x \rangle_q^{\lat} \,.
\label{bare_lat}
\end{eqnarray}
Together with the change to a scheme (here taken as $\overline{MS}$)
\begin{eqnarray}
   \left( \begin{array}{c}
             \langle x \rangle_g^{\msbar}(\mu) \\
             \sum_q \langle x \rangle_q^{\msbar}(\mu) \\
          \end{array}
   \right) 
   = \left( \begin{array}{cc}
               Z_{\bare\,gg}^{\msbar}(\mu)    & 1 - Z_{\bare\,qq}^{\msbar}(\mu) \\
               1 - Z_{\bare\,gg}^{\msbar}(\mu)&   Z_{\bare\,qq}^{\msbar}(\mu)   \\
           \end{array}
     \right) \, \left( \begin{array}{c}
                          \langle x \rangle_g^{\bare} \\
                          \sum_q \langle x \rangle_q^{\bare} \\
                       \end{array}
                \right) \,,
\end{eqnarray}
this completes the renormalisation procedure. As we are considering
quenched QCD only there is a simplification as $Z_{\bare\,gg}^{\msbar} = 1$,
\begin{eqnarray}
   \langle x \rangle_g^{\msbar}(\mu) 
      &=& \langle x \rangle_g^{\bare} 
              + [1 - Z_{\bare\,qq}^{\msbar}(\mu)]
                       \sum_q \langle x \rangle_q^{\bare}
                                                         \nonumber \\ 
   \langle x \rangle_q^{\msbar}(\mu) 
      &=& Z_{\bare\,qq}^{\msbar}(\mu) \langle x \rangle_q^{\bare} \,,
\label{xg_renorm_x_bare}
\end{eqnarray}
($Z_{\bare\,qq}^{\msbar}(\mu)$ is common for all the quarks).
We thus need to determine $Z_g$, $Z_q$ and $Z_{\bare\,qq}^{\msbar}(\mu)$.
We can find  $Z_g$ by following~\cite{michael96a} in considering an
alternative interpretation of the action~(\ref{S_lambda}).
We motivated this action by adding a multiple of the gluon $x$ operator
to the standard action, but we could also write the action as
\begin{eqnarray}
   S = \third\beta_t \sum_i\,\mbox{Re}\,\mbox{tr}_c
               \left[1 - U_{i4}^{\plaq}(\vec{x},\tau) \right]
       + \third\beta_s \sum_{i<j}\, \mbox{Re}\,\mbox{tr}_c
               \left[1 - U_{ij}^{\plaq}(\vec{x},\tau) \right] \,.
\label{S_asym}
\end{eqnarray}
which is the standard way of writing a gluon action on an
anisotropic asymmetric lattice, with differing spatial and temporal
lattice spacings, $a_s \ne a_t$. This action has been studied
in detail, in particular the way in which the anisotropy $\xi = a_s/a_t$
depends on $\beta_s$ and $\beta_t$ is known both perturbatively
and non-perturbatively \cite{engels99a}. At tree-level the anisotropy
is given by $\xi_{tree}^2 = \beta_t / \beta_s$.
$Z_g$ can be found by comparing the anisotropy actually produced
by splitting $\beta_s$ and $\beta_t$ with this tree-level value.
The result is $Z_g = 1 - \frac{g^2}{2} ( c_\sigma - c_\tau )$
where the anisotropy coefficients $c_\sigma$ and $c_\tau$ are defined
in \cite{engels99a}. Using the perturbative values
for $c_{\sigma, \tau}$~\cite{karsch82a} yields $Z_g = 1 - 0.16677 g^2 + \cdots$
as the $1$-loop perturbative $Z_g$. In \cite{meyer07b} this result
was combined with non-perturbative determinations of $c_{\sigma, \tau}$,
\cite{engels99a}, to give a Pad\'e expression
\begin{eqnarray}
   Z_g = { 1 - 1.0225g^2 + 0.1305g^4 \over 1 - 0.8557g^2} \,,
         \qquad \beta \ge 5.7 \,,
\label{Zg_pade}
\end{eqnarray}
(with an error of $\sim 1\%$). So for $\beta = 6.0$ this gives $Z_g = 0.748$.

\begin{figure}[htb]
   \begin{center}
      \includegraphics[width=10.00cm]
         {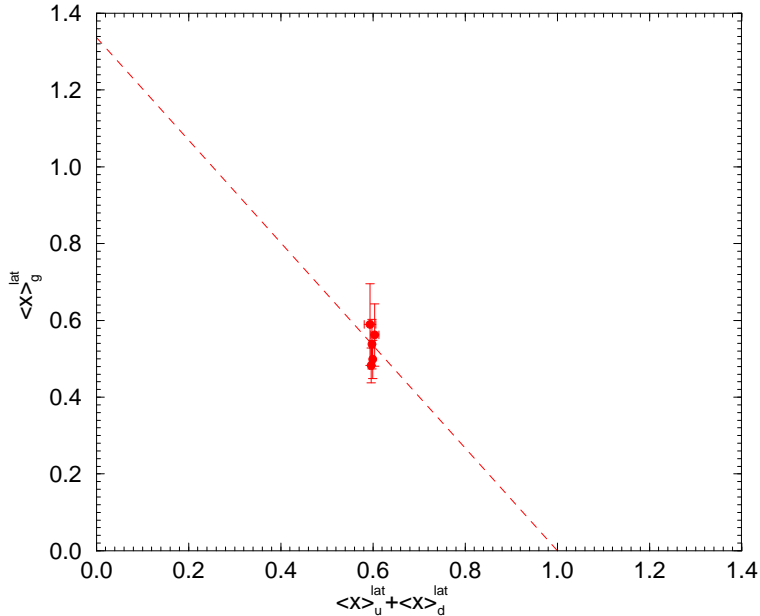}
   \end{center}
\caption{$\langle x \rangle_u^{\lat} + \langle x \rangle_d^{\lat}$
         against $\langle x \rangle_g^{\lat}$ for the five $\kappa$
         values, together with the line $y = (1-x)/0.748$.}
\label{xf_n_lat_sum_xg_n_lat}
\end{figure}

To estimate $Z_q$ we use the results for $\langle x \rangle_g^{\lat}$
from Table~\ref{xg_lat_res} together with those for
$\langle x \rangle_u^{\lat}$, $\langle x \rangle_d^{\lat}$
from \cite{gockeler04a} (i.e.\ $v_{2b}$) together with
eqs.~(\ref{bare_sum_rule}) and (\ref{bare_lat}).
In Fig.~\ref{xf_n_lat_sum_xg_n_lat}
we plot%
\footnote{The total contribution to $\langle x \rangle_q$ from sea quarks has
the form $N_f \times \mbox{(disconnected term)}$. So, even though
the  disconnected loop term is itself non-zero, we do not need
to consider it because its coefficient vanishes if we work
consistently in the quenched approximation.}
$\langle x \rangle_u^{\lat} + \langle x \rangle_d^{\lat}$
against $\langle x \rangle_g^{\lat}$. From eq.~(\ref{bare_sum_rule})
we would expect that the $y$-intercept is given by $1/Z_g$ and
the $x$-intercept is given by $1/Z_q$. At present we do not
have enough results for a determination, so we shall just
check for consistency by fixing the $y$-intercept as $1/0.748$
and the $x$-intercept as $1$, \cite{meyer07a}. This gives
consistency so we shall adopt here $Z_q = 1$ together with a
$10\%$ error.
 
Also from \cite{gockeler04a}, we have for $\mu = 2\,\mbox{GeV}$,
\begin{eqnarray}
   Z^{\msbar}_{bare\,qq}(\mu=2\,\mbox{GeV})Z_q 
      &=& Z^{\rgi}_{v_{2b}}\times[\Delta Z^{\msbar}_{v_2}(\mu=2\,\mbox{GeV})]^{-1}
                                                         \nonumber \\ 
      &=& 1.45 \times 0.732(9) = 1.06(1) \,,
\label{Zmsbar_qq}
\end{eqnarray}
where the second equation uses the notation of that article
(the non-perturbative $RI-MOM$ scheme is converted to an $RGI$ form
and then back to the $\overline{MS}$ scheme). Further values of 
$\Delta Z^{\msbar}_{v_2}(\mu)$ are also given in \cite{gockeler04a}.
With $Z_q$ this then gives $Z^{\msbar}_{bare\,qq}$.


\section{Results and conclusion}
\label{results}

\begin{figure}[htb]
   \begin{center}
      \includegraphics[width=10.00cm]
         {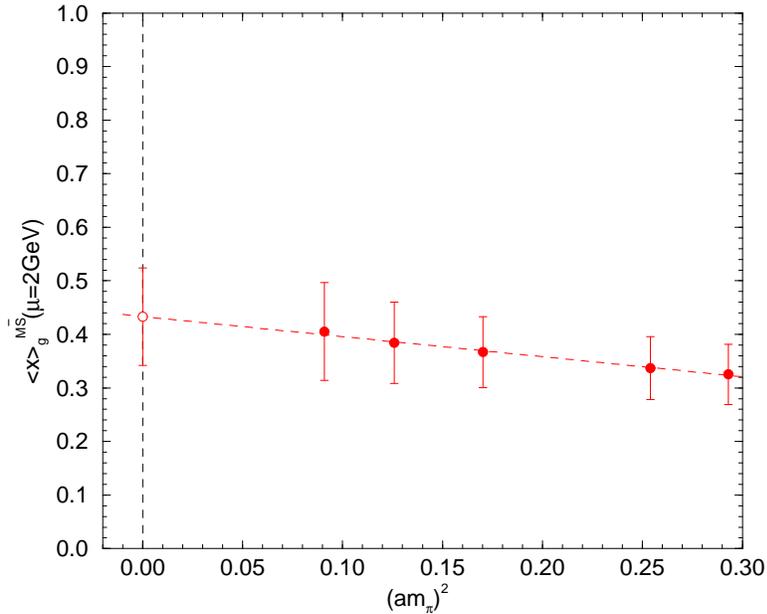}
   \end{center}
\caption{$\langle x \rangle_g^{\msbar}(\mu=2\,\mbox{GeV})$ versus
         $(am_\pi)^2$ for the five $\kappa$ values,
         together with a linear chiral extrapolation.}
\label{x_gluon_n_renorm}
\end{figure}

We are now in a position to determine
$\langle x \rangle_g^{\msbar}(\mu=2\,\mbox{GeV})$. Using the first equation
in eq.~(\ref{xg_renorm_x_bare}) together with eq.~(\ref{Zg_pade})
(evaluated at $\beta = 6.0$) and eq.~(\ref{Zmsbar_qq}) gives
$\langle x \rangle_g^{\msbar}(\mu=2\,\mbox{GeV})$. In 
Fig.~\ref{x_gluon_n_renorm} we plot using eq.~(\ref{xg_renorm_x_bare}),
$\langle x \rangle_g^{\msbar}(\mu=2\,\mbox{GeV})$ versus $(am_\pi)^2$.
This gives a value for $\langle x \rangle_g^{\msbar}(\mu=2\,\mbox{GeV})$ 
of
\begin{eqnarray}
   \langle x \rangle_g^{\msbar}(\mu=2\,\mbox{GeV}) = 0.43(7)(5) \,,
\end{eqnarray}
as our final result, where the first error is in the determination of
$\langle x \rangle_g^{\lat}$ and the second is due to the renomalisation
procedure. This is a significant improvement of our
previous estimate $0.53(23)$ based on generating $O(5000)$
configurations, \cite{gockeler96b} (with error given just for
$\langle x\rangle_g^{\lat}$). 

Direct measurements of gluonic expectation values are
notoriously plagued by noise problems, because the gluons
are bosonic fields. We have seen here that a cheaper alternative,
modifying the gluon action and using the Feynman-Hellmann
theorem to find expectation values from mass measurements,
works well. Here we have performed a test calculation
in the quenched case. The method is a generalisation of that
used to determine the sigma term (see e.g.\ \cite{horsley11a}
and references therein), $\beta$-function, e.g.\ \cite{bali95a},
or singlet terms, e.g.\ \cite{detmold04a}. It is clearly interesting
to repeat this with dynamical fermions.


\section*{Acknowledgements}
\label{acknowledgements}


The numerical calculations were performed on the SGI ICE 8200 at HLRN
(Berlin--Hannover, Germany). This investigation has been supported partly
by the DFG under contract SFB/TR 55 (Hadron Physics from Lattice QCD)
and by the EU grant 283286 (Hadron Physics3). RM is supported by
the EU grant 238353 (ITN STRONGnet) and JMZ by the Australian
Research Council grant FT100100005. We thank all funding agencies.
We would also like to thank W. Bietenholz for a careful reading of
the manuscript and V.~M. Braun and M. G\"ockeler for discussions on
renormalisation.




\begin{thebibliography}{99}

\bibitem{renner10a}
    D.~B. Renner,
    \emph{PoS} (LAT2009) 018, {\tt arXiv:1002.0925}.

\bibitem{hagler09a}
    Ph. H\"agler,
    \emph{Phys.\ Rept.\ } \underline{490} (2010) 49,
    [{\tt arXiv:0912.5483[hep-lat]}].

\bibitem{best97a}
   C. Best, M. G\"ockeler, R. Horsley, E.-M. Ilgenfritz, H. Perlt,
   P. Rakow, A. Sch\"afer, G. Schierholz, A. Schiller and S. Schramm,
   \emph{Phys.\ Rev.\ } \underline{D56} (1997) 2743,
   [{\tt arXiv:hep-lat/9703014}].

\bibitem{guagnelli04a}
   M. Guagnelli, K. Jansen, F. Palombi, R. Petronzio, A. Shindler,
   and I. Wetzorke,
   \emph{Eur.\ Phys.\ J.\ } \underline{C40} (2005) 69,
   [{\tt arXiv:hep-lat/0405027}].

\bibitem{gockeler96b}
    M. G\"ockeler, R. Horsley, E.-M. Ilgenfritz, H. Oelrich, H. Perlt,
    P.~E.~L. Rakow, G. Schierholz, A. Schiller and P. Stephenson,
    \emph{Nucl.\ Phys.\ Proc.\ Suppl.\ } \underline{53} (1997) 324,
    {\tt arXiv:hep-lat/9608017}.

\bibitem{meyer07a}
   H.~B. Meyer and J.~W. Negele,
   \emph{Phys.\ Rev.\ } \underline{D76} (2008) 037501,
   [{\tt arXiv:0707.3225[hep-lat]}].

\bibitem{liu12a}
   K.~F. Liu, M. Deka, T. Doi, Y.~B. Yang, B. Chakraborty, Y. Chen,
   S.~J. Dong, T. Draper, M. Gong, H.~W. Lin, D. Mankame, N. Mathur
   and T. Streuer,
   \emph{PoS} (Lattice 2011) 164, {\tt arXiv:1203.6388}.

\bibitem{edwards04a}
   R.~G. Edwards and B. Jo{\'o},
   \emph{Nucl.\ Phys.\ Proc.\ Suppl.\ } \underline{140} (2005) 832,
   {\tt arXiv:hep-lat/0409003}.

\bibitem{meyer07b}
   H.~B. Meyer,
   \emph{Phys.\ Rev.\ } \underline{D76} (2007) 101701,
   [{\tt arXiv:0704.1801[hep-lat]}].

\bibitem{michael96a}
   C. Michael,
   \emph{Phys.\ Rev.\ } \underline{D53} (1996) 4102,
   [{\tt arXiv:hep-lat/9504016}].

\bibitem{engels99a}
   J. Engels, F. Karsch and T. Scheideler,
   \emph{Nucl.\ Phys.\ } \underline{B564} (2000) 303,
   [{\tt arXiv:hep-lat/9905002}].

\bibitem{karsch82a}
   F. Karsch, 
   \emph{Nucl.\ Phys.\ } \underline{B205} (1982) 285.

\bibitem{gockeler04a}
    M. G\"ockeler, R. Horsley, D. Pleiter, P.~E.~L. Rakow and G. Schierholz,
    \emph{Phys.\ Rev.\ } \underline{D71} (2005) 114511,
    [{\tt arXiv:hep-ph/0410187}].

\bibitem{horsley11a}
   R. Horsley, Y. Nakamura, H. Perlt, D. Pleiter, P.~E.~L. Rakow,
   G. Schierholz, A. Schiller, H. St\"uben, F. Winter and J.~M. Zanotti,
   \emph{Phys.\ Rev.\ } \underline{D85} (2012) 034506,
   [{\tt arXiv:1110.4971[hep-lat]}].

\bibitem{bali95a}
   G.~S. Bali, Ch. Schlichter and K. Schilling,
   \emph{Phys.\ Lett.\ } \underline{B363} (1995) 196,
   [{\tt arXiv:hep-lat/9508027}].

\bibitem{detmold04a}
   W. Detmold,
   \emph{Phys.\ Rev.\ } \underline{D71} (2005) 054506,
   [{\tt arXiv:hep-lat/0410011}].

\end{thebibliography}
\end{document}